\documentclass[12pt]{book}

\usepackage[dvips]{graphicx,color}
\usepackage{makeidx,tsukuba}

\makeauthorindex
\makeindex

\begin{document}

\BookTitle{\itshape The 28th International Cosmic Ray Conference}
\CopyRight{\copyright 2003 by Universal Academy Press, Inc.}
\pagenumbering{arabic}

\chapter{
First-order Fermi Particle Acceleration at Relativistic Shock Waves with
a 'Realistic' Magnetic Field Turbulence Model}

\author{%
%
%
Jacek Niemiec,$^1$ and Micha\l{} Ostrowski$^2$ \\
{\it (1) Institute of Nuclear Physics, ul. Radzikowskiego 152, 31-342 Krak\'ow,
 Poland\\
(2) Astronomical Observatory, Jagiellonian University, ul. Orla 171, 30-244\\ 
 Krak\'ow, Poland} \\
}

\section*{Abstract}
First-order Fermi acceleration process at a relativistic shock wave
is investigated by means of Monte Carlo simulations involving numerical
integration of particle equations of motion in a turbulent magnetic
field near the shock. In comparison to earlier studies a few 'realistic'
features of the magnetic field structure are included. The upstream field
consists of a mean field component inclined at some angle to the shock
normal and finite-amplitude perturbations imposed upon it. The
perturbations are assumed to be static in the local plasma rest frame.
We apply an analytic model for the turbulence with a flat or a Kolmogorov
spectrum within a finite (wide) wave vector range. The magnetic field is
continuous across the shock --- the downstream field structure is derived
from the upstream one from the hydrodynamical shock jump conditions. We
present and discuss the obtained particle spectra and angular
distributions at mildly relativistic sub- and superluminal shocks. We
show that particle spectra diverge from a simple power-law, an exact
shape of the spectrum depends on both an amplitude of the magnetic field
perturbations and the considered wave power spectrum.

\section{Introduction}
At a relativistic shock wave the bulk velocity of the flow is comparable to 
particle velocity. This leads to anisotropy of particle
angular distribution which can substantially influence the process of particle
acceleration. In contrast to the nonrelativistic case, the particle power-law 
spectral
indices depend on the conditions at the shock, including the spectrum and
amplitude of magnetic field perturbations and the mean field inclination to the
shock normal [1-3, 5-10].

In the case of weakly perturbed magnetic field the acceleration process can be
investigated via analytical methods [5-7]. However, if finite-amplitude MHD 
waves are 
present in the medium these approaches are no longer valid and numerical methods 
have to be used. The particle acceleration studies so far applied very simple
models for numerical modeling of the perturbed magnetic field structure [1-3, 9,
10].
The purpose of the present work is to simulate the first-order Fermi
acceleration process at mildly relativistic shock waves propagating  in  
more realistic perturbed magnetic fields, taking into account a wide wave vector
range turbulence with the power-law spectrum and continuity of the magnetic 
field across the shock, involving the respective matching conditions at the
shock.

Below the upstream (downstream) quantities are labeled with 
the index `1' (`2').

\section{Simulations}
In the simulations trajectories of ultrarelativistic test particles are derived
by integrating their equations of motion in the perturbed magnetic field 
[for details see: Niemiec, Ostrowski, in preparation].
We consider a relativistic planar shock wave propagating in rarefied
electron-proton plasma.  
Upstream of the shock the field
consists of the uniform component, $B_{0,1}$, inclined at some angle 
$\psi_1$ to the shock normal 
and finite-amplitude perturbations imposed upon it. The perturbations are 
modeled 
as a superposition of 294 sinusoidal static waves of finite amplitudes [cf. 10]
which have either a flat $(F(k)\sim k^{-1})$ or a Kolmogorov 
$(F(k)\sim k^{-5/3})$ wave power spectrum in the (wide) wave vector range 
$(k_{min}, k_{max})$ and $k_{max}/k_{min}=10^5$.
The shock moves with the velocity 
$u_1$ with respect to the upstream plasma. The downstream flow 
velocity $u_2$ and the magnetic field structure are obtained from the
hydrodynamic shock jump conditions, so that the field is continuous across the
shock. Derivation of the shock compression ratio  
defined  in the shock rest frame as $R = u_1/u_2$, is
based on the approximate formulae derived in Ref. [5]. 
We consider the acceleration process in the particle energy range
where radiative (or other) losses can be neglected.

\begin{figure}[ht]
\includegraphics[scale=0.85]{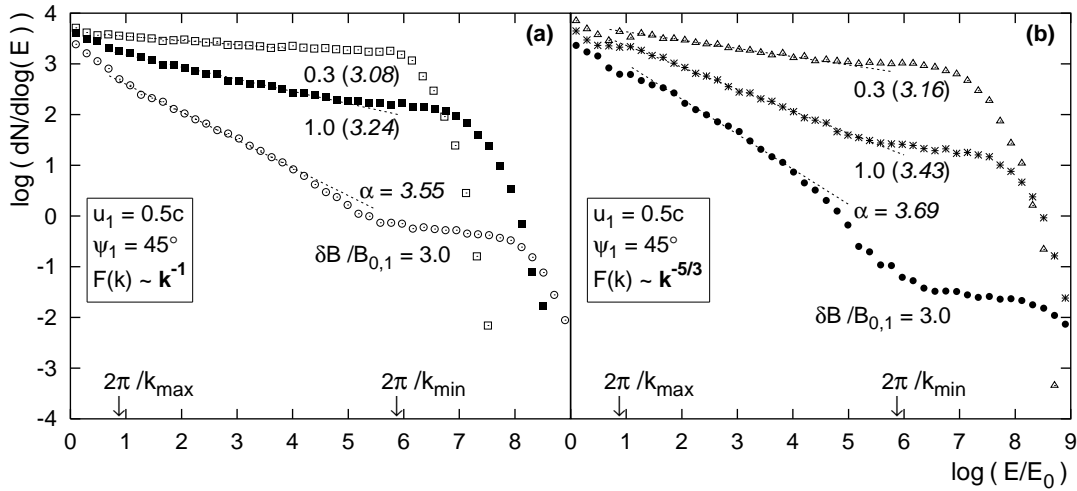} 
\caption{ Accelerated particle spectra at the subluminal shock wave
($u_1=0.5c$, $\psi_1=45^o$ and $u_{B,1}=0.71c$)  for (a)
the flat ($F(k) \propto k^{-1}$) and (b) the Kolmogorov ($F(k) \propto k^{-5/3}$)
wave spectrum of magnetic field perturbations. 
The upstream perturbation amplitude $\delta B / B_{0,1}$
is given near the respective results. Linear fits to the power-law parts of the 
spectra are presented and values of the phase space distribution function spectral
indices $\alpha$ are given. Particles of energies
in the range indicated by arrows can effectively interact with the magnetic field
inhomogeneities ($k_{min} < k_{res} < k_{max}$).} 
\end{figure}

\section{Results}
In Fig. 1 we present particle spectra for the oblique subluminal 
($u_{B,1}\equiv u_1/\cos \psi_1 < c$) shock wave
with  $u_1 = 0.5c$ and $\psi_1 = 45^o$. The shock
velocity along the mean magnetic field is then $u_{B,1} = 0.71c$ and the
shock compression ratio is $R = 5.11$. 
The particle spectra are measured at the shock for three different magnetic
field perturbation amplitudes and the flat (Fig. 1a) and the Kolmogorov 
(Fig. 1b) wave power spectra.
The following features are visible in the spectra:\\
--- the particle spectra diverge from a power-law in the full energy range;\\ 
--- before the spectrum cut-off a harder spectral component can appear;\\
--- the exact shape of the spectrum  depends on both the amplitude of the 
mag- \hspace*{0.55cm}netic  field perturbations and the wave power spectrum.\\
One may note that 
a power-law part of the particle spectrum steepens with increasing amplitude of
the field perturbations, more for the Kolmogorov perturbations. 
The obtained
spectral indices are consistent with previous numerical calculations [9, 10] and 
the analytic results obtained
in the limit of small perturbations [6].

The non power-law character of the obtained particle spectra results from the 
limited dynamic range of magnetic field perturbations. In the energy range
where the approximately
power-law spectrum forms particles can be effectively scattered by the magnetic
field inhomogeneities. The character of the spectrum changes at highest particle 
energies where $k_{res}\leq k_{min}$ and
particles are only weakly scattered. Then the anisotropically distributed
upstream particles can effectively reflect from the region of compressed
magnetic field downstream of the shock leading to the spectrum flattening
[cf. 9]. 
The cut-off in the spectrum is formed mainly
due to very weakly scattered particles escaping from the shock to the introduced
upstream free escape boundary.

The spectra obtained for superluminal shocks with $u_{B,1} \approx 2$ are 
presented in Fig. 3. For the low amplitude turbulence ($\delta B / B = 0.3$) 
we approximately reproduce results
of Ref. [4], with a `super-adiabatic' compression of injected 
particles, but hardly any power-law spectral tail. At larger turbulence 
amplitudes power-law sections in the spectra are produced again, but the 
steepening and the cut-off occur at lower
energies in comparison to the subluminal shocks (Fig. 1).  
\begin{figure}[ht]
\includegraphics[scale=0.85]{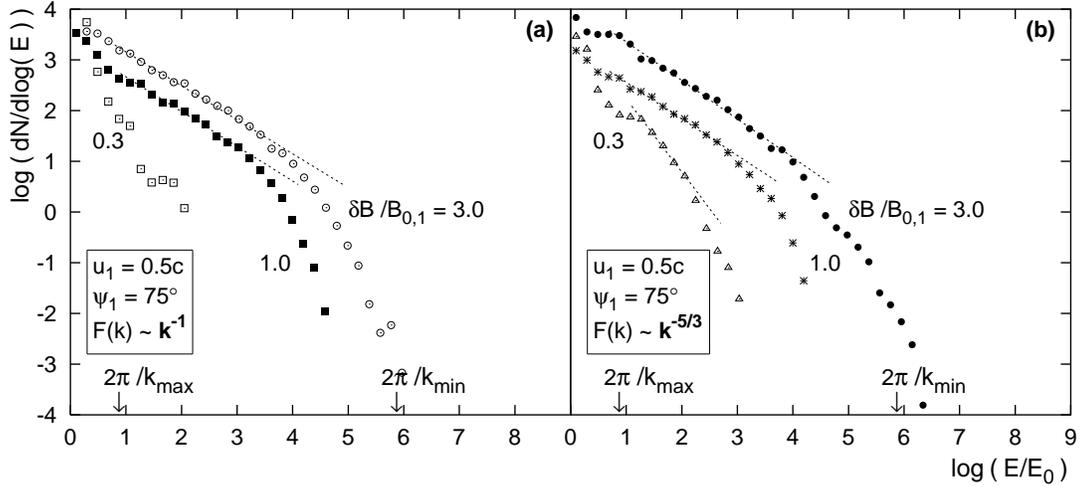} 
\caption{Accelerated particle spectra at the superluminal shock ($u_{B,1}=1.93c$) 
for (a) the flat and (b) the Kolmogorov  
spectra of magnetic field perturbations.} 
\end{figure}
\section{Summary}
The present work is intended to study  the first-order
Fermi acceleration process acting at relativistic shocks.  
In comparison to the previous work we
include a few `realistic' features of the considered turbulence.
We show that the spectrum can deviate from the usually considered power-law 
form and we demonstrate variation of the particle spectral index for its
power-law section with the spectrum and
amplitude of turbulence and the mean field inclination.

The work was supported by the Polish State 
Committee for Scientific Research in 2002-2004 as a research project 2 P03D 008
23 (JN) and in 2002-2005 as a solicitated research project 
PBZ-KBN-054/P03/2001 (M0).

\section*{References}
\noindent
1. Ballard K.R., Heavens A. 1992, MNRAS 259, 89\\
2. Bednarz J., Ostrowski M. 1996, MNRAS 283, 447\\
3. Bednarz J., Ostrowski M. 1998, Phys. Rev.Lett. 80, 3911\\
4. Begelman M.C., Kirk J.G. 1990, ApJ 353, 66\\ 
5. Heavens A., Drury L'O.C. 1988, MNRAS 235, 997\\
6. Kirk J.G., Heavens A. 1989,  MNRAS 239, 995\\
7. Kirk J.G., Schneider P. 1987, ApJ 315, 425\\
8. Kirk J.G., Schneider P. 1987, ApJ 322, 256\\
9. Ostrowski M. 1991, MNRAS 249, 551\\
10. Ostrowski M. 1993, MNRAS 264, 248\\
\endofpaper
\end{document}